\documentclass[twocolumn,showpacs,preprintnumbers,amssymb,nofootinbib]{revtex4}

\usepackage{graphicx}
\usepackage{bm} 
\def\be{\begin{equation}}
\def\ee{\end{equation}}
\def\beq{\begin{eqnarray}}
\def\eeq{\end{eqnarray}}

\def\bes{\begin{eqnarray}}
\def\ees{\end{eqnarray}}

\begin{document}

\title{New instability for rotating black branes and strings}

\author{Vitor Cardoso}
\email{vcardoso@wugrav.wustl.edu} \affiliation{McDonnell Center for
the Space Sciences, Department of Physics, Washington University,
St. Louis, Missouri 63130, USA \\ and \\ Centro de F\'{\i}sica
Computacional, Universidade de Coimbra, P-3004-516 Coimbra,
Portugal}

\author{Jos\'e P. S. Lemos}
\email{lemos@fisica.ist.utl.pt} \affiliation{Centro Multidisciplinar
de Astrof\'{\i}sica - CENTRA, Departamento de F\'{\i}sica, Instituto
Superior T\'ecnico, Universidade T\'ecnica de Lisboa, Av. Rovisco
Pais 1, 1049-001 Lisboa, Portugal}

\date{\today}

\begin{abstract}
The evolution of small perturbations around rotating black branes
and strings, which are low energy solutions of string theory, are
investigated. For simplicity, we concentrate on the Kerr solution
times transverse flat extra dimensions, possibly compactified, but
one can also treat other branes composed of any rotating black hole
and extra transverse dimensions, as well as analogue black hole
models and rotating bodies in fluid mechanics systems. It is shown
that such a rotating black brane is unstable against any massless
(scalar, vectorial, tensorial or other) field perturbation for a
wide range of wavelengths and frequencies in the transverse
dimensions.  Since it holds for any massless field it can be
considered, in this sense, a stronger instability than the one
studied by Gregory and Laflamme. Accordingly, it has also a totally
different physical origin. The perturbations can be stabilized if
the extra dimensions are compactified to a length smaller than the
minimum wavelength for which the instability settles in, resembling
in this connection the Gregory-Laflamme case. Likewise, this
instability will have no effect for astrophysical black holes.
However, in the large extra dimensions scenario, where TeV scale
black holes can be produced, this instability should be important.
It seems plausible that the endpoint of this instability is a
static, or very slowly rotating, black brane and some outgoing
radiation at infinity.
\end{abstract}

\maketitle
\newpage
\section{Introduction}
A consistent theory of quantum gravity, such as string theory, seems
to require the existence of higher dimensions, which in order not to
contradict observational evidence must be compactified on small
scales. String theory has made some important progress in explaining
the entropy of certain black holes by counting microscopic degrees
of freedom, and thus scenarios with extra, compactified dimensions
must be taken seriously. In turn, some of the most interesting
objects to be studied within string theory are those that possess
event horizons, possibly extended in the compact extra dimensions.

In d spacetime dimensions, an event horizon can be topologically a
sphere $S^{{\rm d}-2}$, but when extended to $\rm p$ extra
dimensions, it could naturally have topologies either $S^{{\rm
d}-2+\rm p}$ in which case it is a higher dimensional black hole, or
$S^{{\rm d}-2}\times R ^{\,{\rm p}}$ being then a black p-brane or a
black string in the ${\rm p}=1$ case \cite{tangmyersperryhorosrom}.
The $R^{\,{\rm p}}$ topology of the transverse dimensions can be
compactified giving flat toroidal topological spaces $T^{\,{\rm
p}}$.

One of the first steps towards understanding these extended higher
dimensional solutions is to investigate their classical stability
against small perturbations. If a solution is unstable, then it most
certainly will not be found in nature (unless the instability is
secular) and the solution looses most of its power. Of course the
next question one must ask is what is the final stable result of
such instability.  Now, in d-spacetime dimensions, the
Schwarzschild-Tangherlini geometry is stable against all kinds of
perturbations, massive or massless
\cite{regwheelkodishicardosolemosetal1}. On the other hand, quite
surprisingly, Gregory and Laflamme \cite{gl} showed that this is not
the case for higher dimensional black branes and black strings.
These objects are unstable. The kind of black p-branes originally
studied in \cite{gl} were solutions of ten dimensional low energy
string theory with metric of the form \be ds^2=ds^2_{\rm
Schw}+dx^idx_i\,, \ee where $ds^2_{\rm Schw}$ stands for the
d-dimensional Schwarzschild-Tangherlini line element, the $x^i$ are
the coordinates of the compact dimensions, and $i$ runs from $1$ to
$p$. The total dimension of the spacetime D obeys ${\rm D}={\rm
d}+{\rm p}$. In string or supergravity theories one takes ${\rm
D}=10$ or ${\rm D}=11$, but for genericity one can leave it as a
free natural parameter.  In \cite{gl} it was shown that even though
scalar and vector perturbations of the black brane stay bounded in
time, the tensorial sector of gravitational perturbations displays
an instability, with the perturbations growing exponentially with
time, possibly breaking the extended black brane into several
smaller black holes.  The only way around this instability is to
compactify the transverse directions $x^i$ on a scale smaller than
the black hole radius \cite{gl} (for a full discussion of this
instability and its developments see \cite{kol}).

Here we shall show that a related class of metrics, for example,
having the form \be ds^2=ds^2_{\rm
Kerr}+dx^idx_i\,,\label{blackkerr} \ee where the line element
$ds^2_{\rm Kerr}$ stands for the d-dimensional Kerr$-$Myers-Perry
line element, is unstable to scalar perturbations, as well as to
vectorial, gravitational and other type of perturbations. In this
sense it is a stronger instability than the one studied in
\cite{gl}, because it holds for any massless field. Accordingly, it
has also a totally different physical origin, as we will see.

\section{Formulation of the problem and basic equations}
\label{formulation}
\subsection{The background metric}
In four dimensions, there is only one possible rotation axis for a
cylindrically symmetric spacetime, and there is therefore only one
angular momentum parameter.  In higher dimensions there are several
choices for rotation axes and there is a multitude of angular
momentum parameters, each referring to a particular rotation axis.
Here we shall concentrate on the simplest case, for which there is
only one angular momentum parameter, which we shall denote by $a$.
The specific metric we shall be interested in is given in
Boyer-Lindquist-type coordinates by
\begin{eqnarray}
ds^2&=& -{\Delta-a^2\sin^2\theta\over\Sigma}dt^2
-{2a(r^2+a^2-\Delta)\sin^2\theta \over\Sigma}
dtd\varphi \nonumber\\
&&{}+{(r^2+a^2)^2-\Delta a^2 \sin^2\theta\over\Sigma} \sin^2\theta
d\varphi^2
\nonumber\\
&&{} +{\Sigma\over\Delta}dr^2 +{\Sigma}d\theta^2+r^2\cos^2\theta
d\Omega_{n}^2+dx^idx_i, \label{metric}
\end{eqnarray}
where
\begin{eqnarray}
\Sigma&=&r^2+a^2\cos^2\theta,\\
\Delta&=&r^2+a^2-m r^{1- n},
\end{eqnarray}
and $d\Omega_{n}^2$ denotes the standard metric of the unit
$n$-sphere ($n={\rm d}-4$), the $x^i$ are the coordinates of the
compact dimensions, and $i$ runs from $1$ to $p$. This metric
describes a rotating black brane in an asymptotically flat, vacuum
space-time with mass and angular momentum proportional to $\mu$ and
$\mu a$, respectively. Hereafter, $\mu,a>0$ are assumed.

The event horizon, homeomorphic to $S^{2+ n}$, is located at
$r=r_+$, such that $\Delta|_{r=r_+}=0$.  For $n=0$, an event horizon
exists only for $a<\mu/2$. When $n=1$, an event horizon exists only
when $a<\sqrt{\mu}$, and the event horizon shrinks to zero-area in
the extreme limit $a\rightarrow\sqrt{\mu}$. On the other hand, when
$n\ge 2\,,$ $\Delta=0$ has exactly one positive root for arbitrary
$a>0$. This means there is no bound on $a$, and thus there are no
extreme Kerr black branes in higher dimensions.

\subsection{Separation of variables and boundary conditions}
Consider now the evolution of a massless scalar field $\Psi$ in the
background described by (\ref{metric}). The evolution is governed by
the curved space Klein-Gordon equation \be \frac{\partial}{\partial
x^{\mu}} \left(\sqrt{-g}\,g^{\mu \nu}\frac{\partial}{\partial
x^{\nu}}\Psi \right)=0\,, \label{klein} \ee where $g$ is the
determinant of the metric. The metric appearing in (\ref{klein})
should describe the geometry referring to both the black brane and
the scalar field, but if we consider that the amplitude of $\Psi$ is
so small that its contribution to the energy content can be
neglected, then the metric (\ref{metric}) should be a good
approximation to $g_{\mu \nu}$ in (\ref{klein}).  We shall thus work
in this perturbative approach.  It turns out that it is possible to
simplify considerably equation (\ref{klein}) if we separate the
angular variables from the radial and time variables, as is done in
four dimensions \cite{brill}. For higher dimensions we follow
\cite{ida}. In this connection see also \cite{frolovpage} for a
general $4+n$-dimensional Kerr hole with several spin parameters. In
the end our results agree with the results in \cite{frolovpage}, if
we consider only one angular momentum parameter in their equations,
and no extra compactified dimensions.

We consider the ansatz $\phi=e^{i\omega
t-im\varphi+ik_ix^i}R(r)S(\theta)Y(\Omega)$, and substitute this
form in (\ref{klein}), where $Y(\Omega)$ are hyperspherical
harmonics on the $n$-sphere, with eigenvalues given by $-j(j+n-1)$
($j=0,1,2,\cdots$). Then we obtain the separated equations
\begin{eqnarray}
&&{1\over\sin\theta\cos^n\theta}\left({d\over d \theta}
\sin\theta\cos^n\theta{dS\over d \theta}\right)
+\left[a^2(\omega^2-k^2)\cos^2\theta \right.
\nonumber\\
&&{}\left. -m^2\csc^2\theta -j(j+n-1)\sec^2\theta +A\right]S=0,
\label{ang}
\end{eqnarray}
and
\begin{eqnarray}
&&r^{-n}{d\over d r}\left(r^n\Delta{dR\over d r}\right) + \left\{
{\left[\omega(r^2+a^2)-ma\right]^2\over\Delta} \right.
\nonumber\\
&&{}\left. -{j(j+n-1)a^2\over r^2} -\lambda-k^2r^2
\right\}R=0, \label{rad}
\end{eqnarray}
where $A$ is a constant of separation, $\lambda:=A-2m\omega
a+\omega^2 a^2$, and $k^2=\sum k_{i}^2$. Interestingly, note the
important point that Equations (\ref{ang})-(\ref{rad}) are just
those that describe the evolution of a massive scalar field, with
mass $k$, in a d-dimensional Kerr geometry.

The equations (\ref{ang}) and (\ref{rad}) must be supplemented by
appropriate boundary conditions, which are given by
\begin{equation}
R\sim\left\{
\begin{array}{ll}
(r-r_H)^{i\sigma} & {\rm as}\ r\rightarrow r_H \,, \\
r^{-(n+2)/2} {\rm e}^{-i{\sqrt{\omega^2-k^2} r}} & {\rm as}\
r\rightarrow \infty\,.
\end{array}
\right. \label{bound2}
\end{equation}
where
\begin{equation}
\sigma:={\left[(r_H^2+a^2)\omega-ma\right]r_H \over
(n-1)(r_H^2+a^2)+2r_H^2} \,,
\end{equation}
has been determined by the asymptotic behavior of the
Eq.~(\ref{rad}). In other words, the waves must be purely ingoing at
the horizon and purely outgoing at the infinity.  For assigned
values of the rotational parameter $a$ and of the angular indices
$l\,,j\,,m$ there is a discrete (and infinite) set of frequencies
called quasinormal frequencies, QN frequencies or $\omega_{QN}$,
satisfying the wave equation (\ref{rad}) with the boundary
conditions just specified by Eq. (\ref{bound2}).

\section{The instability timescale}
\label{Results}

Now, when ${\rm d}=4$, analysis of the perturbations of a massive
scalar field on the Kerr geometry has been done
\cite{massivescalar}. Indeed, equations (\ref{ang})-(\ref{rad}) have
been analytically solved by Zouros and Eardley  in the limit of
large $k^2$ and by Detweiler in the limit of small $k^2$
\cite{massivescalar}. Detweiler's results have recently been
confirmed numerically by Furuhashi and Nambu \cite{massivescalar}.
Detweiler \cite{massivescalar} shows that for small $k^2$ in
particular $k M<<1$, the geometry is unstable. In fact, he shows
that the characteristic frequencies $\omega$ satisfying the boundary
conditions (\ref{bound2}) are given by \be \omega= \sigma +
i\gamma\,, \ee with \be \sigma ^2=k^2\left ( 1-\left (\frac{k
M}{l+1+n}\right )^2\right )\,,n=0,1,2...\,, \ee and \be
\gamma=C\left (k-\frac{am}{2Mr_+} \right )\,, \ee with
$C=C(l,m,a,M,n)$ a positive constant, as shown in Detweiler
\cite{massivescalar}. So, for $m$ positive and \be
k<\frac{am}{2Mr_+} \ee which is just a superradiance condition, the
mode is then unstable. In particular, the most unstable mode is the
$l=m=1$ mode with an $e$-folding time $\tau $ given by of \be
\frac{\tau}{M}=24\,\frac{1}{k\,a}\,\frac{1}{(kM)^{8}}\,.
\label{insttime} \ee Now, we know their results can be translated
immediately into our black brane geometry. Thus rotating black
branes of the form (\ref{metric}) with ${\rm d}=4$ and p extra
dimensions are unstable, with an $e$-folding time given by
(\ref{insttime}). The results for ${\rm d}>4$ are not available in
the literature, but the instability should be present in these cases
as well.

The only way around this instability is to make the superradiant
factor $k-\frac{am}{2Mr_+}$ positive, which is the same as having
$kM>\frac{am}{2r_+}$. Thus, for sufficiently small $a/r_+$, this is
still inside the range of validity of our approximations, $kM\ll 1$.
To escape this instability one must compactify the transverse
dimensions on a small scale $L$, i.e, $\frac{L}{2\pi M}<\frac{2
r_+}{a}$. Writing out $r_+$ explicitly this gives \be \frac{L}{2\pi
M}<2\,\alpha\,, \ee where $\alpha=M/a+\sqrt{M^2/a^2-1}\sim 2M/a$,
for small $a$. From (\ref{insttime}) one obtains that the timescale
for the instability, in the approximations used is
$\tau/M>24(1/kM)^9$. Since $kM\ll 1$ this is not a specially
efficient instability, and it doesn't even appear to overcome the
Gregory-Laflamme instability timescale. However, the only results
available so far for the Gregory-Laflamme instability refer to
non-rotating objects. It is possible that as one increases the
rotation to large values the instability studied here becomes more
effective (as is apparent from Eq. (\ref{insttime})) and that the
Gregory-Laflamme instability gets less effective.

\section{Conclusions}
\label{Conclusions}
What is the physical interpretation of this instability for rotating
black branes? It is known that the Kerr geometry displays
superradiance \cite{zelmisnerunruhbekenstein}. This means that in a
scattering experiment of a wave with frequency $\omega<m\Omega$ the
scattered wave will have a larger amplitude than the incident wave,
the excess energy being withdrawn from the object's rotational
energy. Here $\Omega$ is the horizon's angular velocity (related to
$a$, $M$ and $r_+$ through $\Omega=a/(2Mr_+)$) and $m$ is again the
azimuthal wave quantum number. Now suppose that one encloses the
rotating black hole inside a spherical mirror. Any initial
perturbation will get successively amplified near the black hole
event horizon and reflected back at the mirror, thus creating an
instability. This is the black hole bomb, as devised in
\cite{teupress} and recently improved in \cite{cardosoetal}. This
instability is caused by the mirror, which is an artificial wall,
but one can devise natural mirrors if one considers massive fields.
In this case, the mass of the field acts effectively as a mirror,
and thus Kerr black holes are unstable against massive field
perturbations \cite{massivescalar}. With this in mind, we expect
that black strings and branes of the form (\ref{blackkerr}) will be
unstable, because the compactified transverse directions work as an
effective mass for the graviton and for the scalar field. In fact,
this simple reasoning implies that any rotating black brane (more
general than the one described by (\ref{blackkerr})) will be
unstable. This would be the case for the rotating black hole in
string theory found by Sen or other rotating black holes
\cite{sencvetic}. Moreover, following Zel'dovich
\cite{zelmisnerunruhbekenstein}, it is known that not only the Kerr
geometry, but any rotating absorbing body for that matter displays
superradiance. Thus, this instability should appear in analogue
black hole models \cite{cardoso}, and in rotating bodies in fluid
mechanics systems.

Here, we have derived the instability timescale for a scalar field,
and not for geometry (metric) perturbations, since Teukolsky's
formalism for higher dimensional rotating objects is not
available. Still, the argument presented above makes it clear that
the instability should be present for metric perturbations as well.
We also expect that the instability will be stronger for metric
modes, because of the following simple reasoning. Superradiance is
the mechanism responsible for this instability, and thus the larger the
superradiant effects, the stronger the instability. Now, we know
that in the Kerr geometry in ${\rm d}=4$
scalar fields have a maximum
superradiant amplification factor of about $2\%$, whereas
gravitational modes have maximum superradiant amplification factor
of about $138 \%$ \cite{ampfact}. Then it is expected
the instability
timescale to be almost two orders of magnitude smaller,
and correspondingly, the instability should
grow much stronger for gravitational modes.
For general black branes, the Gregory-Laflamme
instability seems to be stronger than the one displayed here, but it
is known that certain extremal solutions should not exhibit the
Gregory-Laflamme instability \cite{reall}, whereas the instability
dealt with here should go all the way to extremality. So eventually
it takes over the Gregory-Laflamme instability. Moreover, recent
studies \cite{maartens} seem to indicate that black strings in a
Randall-Sundrum inspired 2-brane model do not exhibit the
Gregory-Laflamme instability.

The endpoint of this rotating instability is not known, and it can never be
predicted with certainty by a linear analysis. However, it seems
plausible to assume that the instability will keep growing until the
energy and angular momentum content of the field approaches that of
the black brane, when back-reaction effects become important. The
rotating brane will then begin to spin down, and gravitational and
scalar radiation goes off to infinity carrying energy and angular
momentum. The system will probably asymptote to a static, or very
slowly rotating, final state consisting of a non-rotating black
p-brane and some outgoing radiation at infinity. But we cannot
discard the other possibility that the horizon fragments.

\centerline{}
\centerline{}
\centerline{}
\section*{Acknowledgements}
We gratefully acknowledge stimulating correspondence on this problem
with Donald Marolf. V. C. acknowledges financial support from
Funda\c c\~ao para a Ci\^encia e Tecnologia (FCT) - Portugal through
grant SFRH/BPD/2003. This work was partially funded by FCT through
project POCTI/FNU/44648/2002.

\end{document}